\documentclass[conference]{IEEEtran}

\usepackage{epsfig} 
\usepackage{graphicx}
\usepackage{floatrow}
\usepackage{amssymb}
\usepackage{amsfonts}
\usepackage{latexsym}
\usepackage[cmex10]{amsmath}
\usepackage{array}
\usepackage{cite} 
\usepackage{import}  
\usepackage[squaren]{SIunits}
\usepackage{grffile}
\usepackage{multirow}
\usepackage{booktabs} 
\usepackage{epstopdf}
\usepackage{import}
\usepackage{color}
\usepackage[utf8]{inputenc}
\usepackage{xcolor}
\usepackage{balance}
\usepackage{subcaption}
\usepackage{url}

\newcolumntype{L}{>{\raggedright\arraybackslash}X}

\def\BibTeX{{\rm B\kern-.05em{\sc i\kern-.025em b}\kern-.08em
		T\kern-.1667em\lower.7ex\hbox{E}\kern-.125emX}}

\newcommand{\me}{\text{e}}

\newcommand{\mj}{\text{j}}

\newcommand{\bm}{\boldsymbol}

\begin{document}

\title{Frame Error Rate Prediction for Non-Stationary Wireless Vehicular  Communication Links}

\author{
\IEEEauthorblockN{Anja Dakić\textsuperscript{}, Benjamin Rainer\textsuperscript{}, Markus Hofer\textsuperscript{}, Thomas Zemen\textsuperscript{}}
\IEEEauthorblockA{
	\\\textsuperscript{}\textit{Center for Digital Safety \& Security}, \textit{Austrian Institute of Technology GmbH}, Vienna, Austria \\
	    anja.dakic@ait.ac.at}
}

\maketitle

\begin{abstract}
    Wireless vehicular communication will increase the safety of road users. The reliability of vehicular communication links is of high importance as links with low reliability may diminish the advantage of having situational traffic information. The goal of our investigation is to obtain a reliable coverage area for non-stationary vehicular scenarios. Therefore we propose a deep neural network (DNN) for predicting the expected frame error rate (FER). The DNN is trained in a supervised fashion, where a time-limited sequence of channel frequency responses has been labeled with its corresponding FER values assuming an underlying wireless communication system, i.e. IEEE 802.11p. For generating the training dataset we use a geometry-based stochastic channel model (GSCM). We obtain the ground truth FER by emulating the time-varying frequency responses using a hardware-in-the-loop setup. Our GSCM provides the propagation path parameters which we use to fix the statistics of the fading process at one point in space for an arbitrary amount of time, enabling accurate FER estimation. Using this dataset we achieve an accuracy of $85\,\%$ of the DNN. We use the trained model to predict the FER for measured time-varying channel transfer functions obtained during a measurement campaign. We compare the predicted output of the DNN to the measured FER on the road and obtain a prediction accuracy of $78\,\%$.
\end{abstract}

\vspace{1mm}
\begin{IEEEkeywords} 
IEEE 802.11p, FER, deep learning, GSCM, hardware-in-the-loop
\end{IEEEkeywords}

\section{Introduction}

Advanced driver-assistance systems (ADAS) shall increase the safety of road users but currently rely on a local view of the traffic situation. Thus, it is missing valuable information that may be perceived by other road users. In vehicle-to-everything (V2X) communication we want to exchange situational traffic information directly between two parties, the transmitter (Tx) and the receiver (Rx). This information has to be timely transmitted using wireless communication. Hence, it is of high interest to know a priori whether the information will be received by other road users in the near vicinity.  

V2X communication scenarios have typically highly dynamic wireless propagation characteristics. Furthermore, the direct propagation path between the Tx and the Rx in urban scenarios is often obstructed by other road users or buildings. These dynamic and harsh radio propagation conditions result in time-varying and non-stationary radio channels \cite{Matz05}, \cite{Bernado14}, \cite{Bernado15}.
Therefore, to achieve a reliable information exchange, we investigate methods to estimate a reliable communication region. We determine the reliability of radio communication channels by the frame error rate (FER).

Most of the currently available literature focuses on methods for estimating and mapping a coverage region based on path loss~\cite{Phill13}. The main idea of these models is to compute the free-space path loss and the path loss due to large scale fading.
Besides the standard models, in recent years machine learning (ML) has been applied for predicting path loss. In \cite{Wen19} ML is used to predict the path loss of wireless communication in an aircraft cabin. The authors in \cite{Wen19} use the channel impulse response obtained from real-word measurements to predict the received power. The random forest algorithm is used to predict the path loss in a vehicular channel in \cite{Panthangi19}. The training dataset consists of the extracted information from the measurements, such as the location of the Tx and the Rx as well as the distance between them, and additionally of a specific integer value differentiating the objects between Tx and Rx. The results show improved accuracy in comparison to traditional long-distance path loss prediction models.

In~\cite{Dak20} we show that the FER does not only depend on the path loss but that the five key channel parameters: (i) received power, (ii) root mean square (RMS) delay spread, (iii) RMS Doppler spread, (iv) Rician $K$-factor, and (v) Doppler shift of line-of-sight (LOS) component, are jointly good candidate features for predicting the FER.
Additionally, the FER is determined by transmission parameters (e.g., frame length, modulation format, transmit power) and the specific Rx architecture (e.g., channel estimation, and decoder of forward error correction code). Considering these dependencies, predicting the FER becomes a complex process, especially if we need to do this directly on the road.

A selected number of packet error rate (PER) models for wireless communications are shown in~\cite{Masnikosa20}. The authors compare different models based on the approximation method, the modulation scheme and coding. All of these models are applied for Rayleigh channels. The results in the literature indicate that linear models fail to accurately predict the FER in urban scenarios and under harsh radio wave propagation conditions.

In~\cite{Jam21} deep learning (DL) is applied to predict the FER for collaborative intelligent radio networks. The prediction problem is defined as a binary classification, where the decision is made whether the frame was decoded successfully on the receiver side. As dataset, they use the frame decoding error, the noise variance, the allocated channel for transmission, the modulation and coding scheme, the bandwidth and the power spectral density.

In this work, we choose a machine learning approach to predict the FER of a vehicular communication link. More precisely we want to predict the corresponding FER class. We select the channel transfer function (CTF) as starting point for our investigation. For time-variant and in general non-stationary vehicular radio channels, we assume that the radio channel is wide-sense stationary and exhibits uncorrelated scattering for a specific spatial region which can also be expressed in terms of a time and frequency window, termed stationarity region. Therefore, the FER prediction is performed per stationarity region. As we discussed above, the FER depends on the communication system, therefore we restrict our investigation without loss of generality to IEEE 802.11p and a specific hardware implementation thereof. We design a deep neural network (DNN) for predicting the FER from CTF samples as input. We use a supervised approach to train the DNN which shall solve a classification task, where each class represents a FER range: $ (10^{-x_{1} },\, 10^{-x_{2} }], \, x_{1},x_{2} \in \mathbb{N}_{0}, \, x_{2} < x_{1}$. 
\subsection*{Scientific Contributions of the Paper}
\begin{itemize} 
	\item We propose a new FER measurement methodology for non-stationary vehicular wireless communication scenarios. The methodology allows to measure the FER of a stationarity region of the wireless communication channel with arbitrary accuracy despite its finite length.
    \item We formulate the prediction of the FER in vehicular scenarios as a classification problem and we create a labeled dataset for supervised learning. For the dataset we employ a geometry-based stochastic channel model (GSCM) and a stochastic channel model.
	\item We train a DNN with the dataset and obtain a good accuracy of approx. $85\%$.
    \item We validate the proposed DNN with measurement data and show that we can predict the FER for the given channel conditions of $78\%$.
\end{itemize}

\section{Measuring frame error rate in non-stationary scenarios}
\label{sec:meas:fer}

The propagation conditions in vehicular wireless channels change fast, but, as mentioned before, for a limited period in time and in frequency we may assume that the channel statistics stay constant. This stationarity region is usually short and~\cite{Bernado12} shows that in urban scenarios, constant channel statistics can be assumed for spatial regions of several wavelengths $\lambda$ or in terms of time for less than $100 \,\text{ms}$, for a velocity of about $ 50\,\text{km/h} \approx 13.8\, \text{m/s}$. This in turn leaves us with a very limited number of frames that can be transmitted under constant channel statistics and hence limits the FER resolution for a single stationarity region. 

However, we want to achieve an accurate prediction of the expected FER and therefore we develop a methodology that allows us to send arbitrarily many frames under the same fading characteristics. For this purpose, we use the propagation path parameters obtained from a geometry-based stochastic channel model. We fix these path parameters at defined points in space, then we extend the time-variant frequency response sequence continuously over time and feed it into a hardware-in-the-loop (HiL) setup \cite{Dak20}. 
\subsubsection{Channel Model}
The time-limited frequency response within a stationarity region is calculated by
\begin{equation}
H[m, k] = \sum_{l=1}^{L_{}} \eta_{l}[m] \me^{-\mj 2 \pi (f_\text{C} + k \Delta f) \tau^{}_{l}[m]},
\label{eq:channel}
\end{equation}
where $\eta_{l}[m] = |\eta_{l}[m]|\me^{\mj 2 \pi \phi_{l}} \in \mathbb{C}$ represents a complex time-variant weighting coefficient, which includes the amplitude $|\eta_{l}[m]|$ and the random starting phase $\phi_{l}$, $\tau_l[m] \in \mathbb{R}^+_0$ the time-variant delay, $f_{\text{C}}$ the carrier frequency, and $ \Delta f $ the subcarrier spacing. The path index $l \in\left\lbrace 1,\ldots, L\right\rbrace $, where $L$ is the number of propagation paths. The time index within one stationarity region is denoted by $m\in\left\lbrace 0,\ldots, M-1\right\rbrace$, where $M = T_{\text{stat}}/T_{\text{s}}$ is the total number of samples in time within a stationarity region. The frequency index is denoted by $k \in \{ -\lfloor \frac{B}{2\Delta f} \rfloor, \ldots, \lfloor \frac{B}{2\Delta f} \rfloor - 1 \}$, where $B$ represents the system bandwidth. The number of subcarriers $N$ within one stationarity region is equal to $B/\Delta f$.  

From this formula, we notice that the frequency response is described by the propagation path parameters, namely the attenuation $\eta_{l}[m]$, and the path delay $\tau_{l}[m]$. Within $T_{\text{stat}}$, we assume an approximately constant amplitude of the propagation paths $|\eta_{l}|$, and a constant relative velocity per propagation path $v_{l}$ between the Tx and Rx. This assumption leads us to a constant Doppler shift, which is defined as 
\begin{equation}
    f_{l} = f_{\text{C}}\frac{v_l}{c_{0}},
\end{equation}
where $c_{0}$ is the speed of light in a vacuum. Furthermore, we model the change in delay by a linear model
    \begin{equation}
    \tau_{l}[m] = \tau_{l}[0] - \frac{v_{l}}{c_{0}} m T_{\text{s}}
    \label{eq:ConstVel}.
\end{equation}
Taking these assumptions into account, and extending $m>M-1$ we can create a time-varying fading process with the statistical properties defined by the propagation path parameters for an arbitrary amount of time. Finally, we can write \eqref{eq:channel} as
\begin{equation}
    H[0, k] = \sum_{l=1}^{L_{}} \eta_{l}[0] \me^{-\mj 2 \pi (f_\text{C} + k \Delta f)( \tau^{}_{l}[0]-\frac{f_{l}}{f_{\text{C}}}mT_{\text{s}})}.
\end{equation}
The next sections will show that this is sufficient for our investigations.

\subsubsection{Hardware-in-the-loop Measurements}
In order to measure the FER, in our work we use the HiL framework presented in~\cite{Dak20}. It contains hardware modems as Tx and Rx, and the geometry-based AIT channel emulator~\cite{Hofer19}. The geometry-based AIT channel emulator uses directly the propagation path parameters to compute the time-variant impulse response in the FPGA for the convolution with the transmitted signal. The emulator can deal with real valued path delay and Doppler shifts using the special properties of discrete prolate spheroidal sequence \cite{Slepian78}.

By fixing the propagation path parameters in one stationarity region we can increase the emulation time according to the required number of frames, keeping the very same propagation statistics during that time. Therefore, by applying only a time average, this methodology enables us to detect the FER which is close to an expected FER value. With this approach, we can evaluate the FER with arbitrary accuracy in non-stationary vehicular scenarios with a short stationarity time.
 
\section{Dataset}

\subsection{Input Data}

For training the DNN presented in Section~\ref{sec:DNN}, we create a labeled dataset consisting of time-limited sequences of frequency responses. We obtain them using the GSCM presented in~\cite{Rainer20} and using the stochastic channel model shown in~\cite{Dak20}.

\subsubsection{Geometry-Based Stochastic Channel Model}
We parameterize our GSCM by a measurement campaign conducted in an urban environment of the inner city of Vienna~\cite{Zel21}. The carrier frequency is $f_{\text{C}}=5.9 \, \text{GHz}$ with a bandwidth of $B = 150.25 \, \text{MHz}$ (subcarrier spacing $\Delta f = 250\,\text{kHz}$), and a snapshot duration $T_{\text{s}} = 500\, \text{$\mu$s}$. The maximum speed of both vehicles is $\approx 11\, \text{m/s}$. 

Here, the propagation path parameters, needed for the methodology presented in Section~\ref{sec:meas:fer}, are obtained from the GSCM. We import the geometric environment information from OpenStreetMap (OSM) and vehicle trajectories from the recorded GPS data. Diffuse scatterers are randomly distributed along the building's walls. Other contributions which we differentiate in our GSCM are static discrete scatterers (SD), mobile discrete scatterers, and LOS between Tx and Rx. Each of these scatterer types is described by the propagation path parameters $\eta_{l}, f_{l} \,\text{and}\, \tau_{l}$. The time-limited frequency responses are calculated using \eqref{eq:channel} by collecting the signal contribution of the different types of scatterers. 

Furthermore, in this paper, we cover the vehicle-to-infrastructure (V2I) scenarios and hence we keep the Tx moving along its trajectory and we place $S \in \mathbb{N}$ points along the road to indicate the Rx positions. During one simulation run, i.e. one scatterer realization, we obtain time-varying frequency responses for all Rx positions. In total we generate $7116$ time-limited frequency response sequences with a length of $T_{\text{stat}}=100\,\text{ms}$.

\begin{figure}[h!]
	\begin{center}
		\includegraphics[scale=0.58]{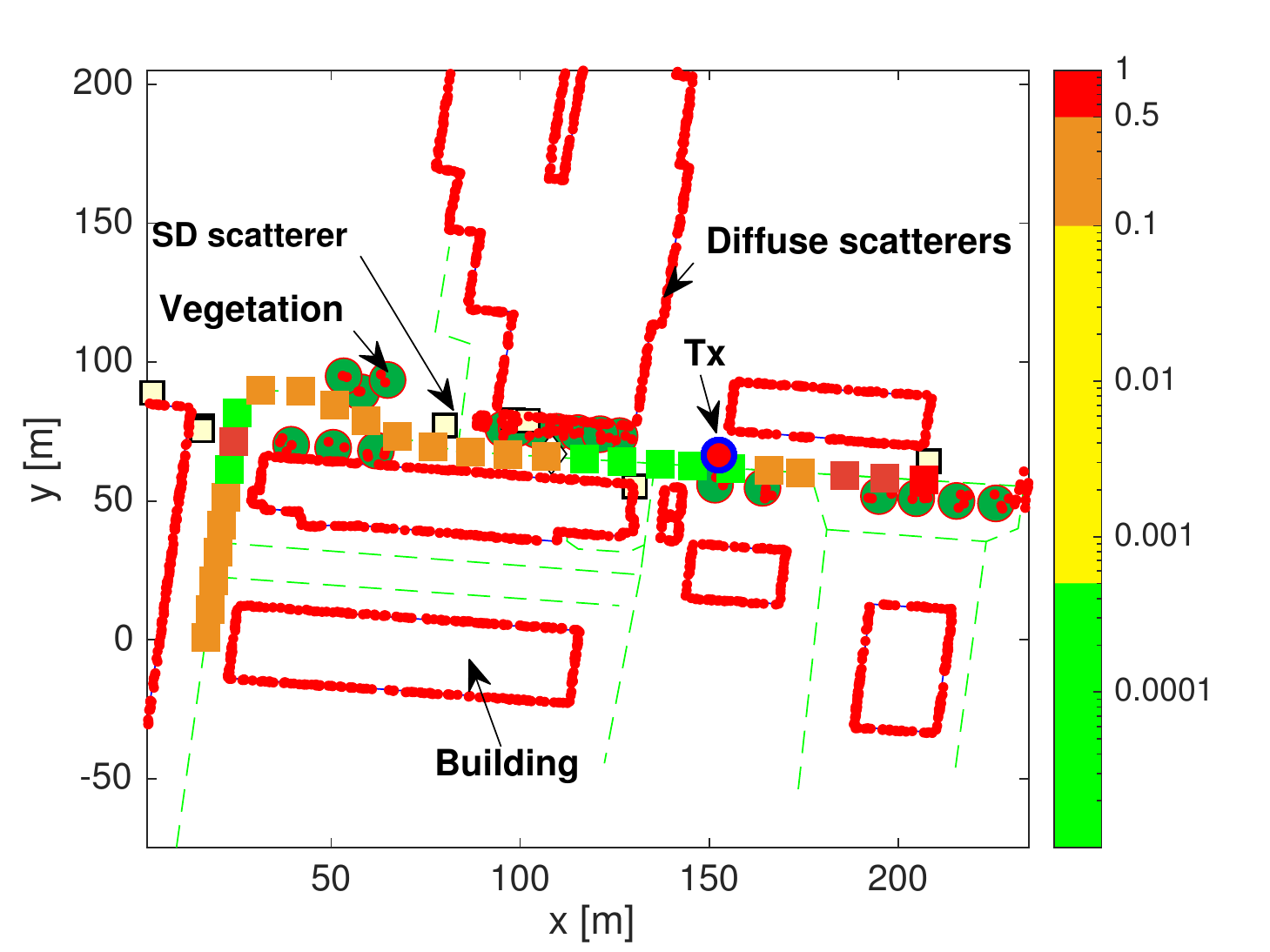}
	\end{center}
	\caption{FER measured at different Rx positions during one stationarity region.}
	\label{fig:FERoneStat} 
\end{figure}

\begin{figure}[h!]
	\begin{center}
		\includegraphics[scale=0.58]{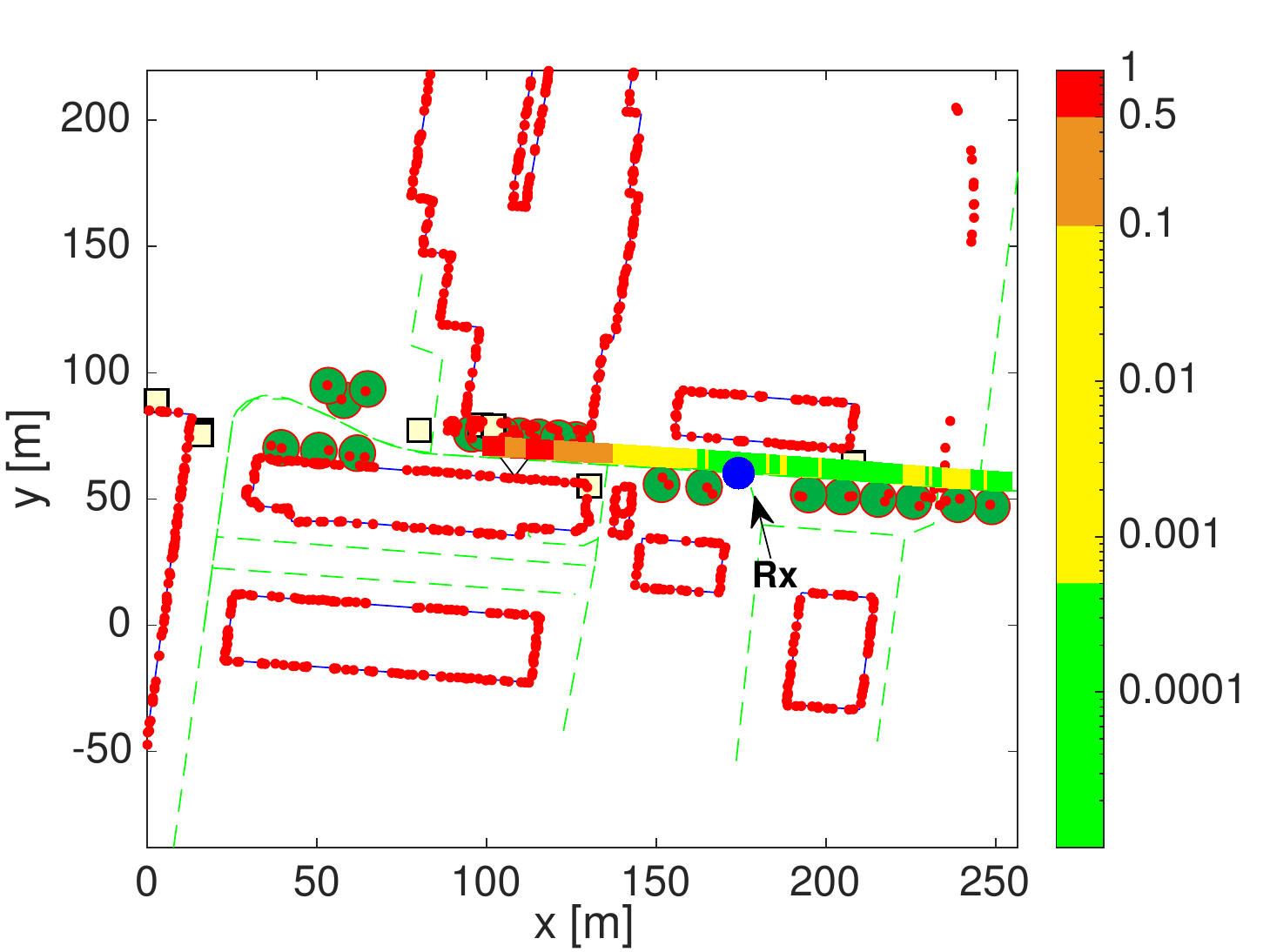}
	\end{center}
	\caption{FER over Tx trajectory for a single Rx position.}
	\label{fig:FERoneRx} 
\end{figure}

In Fig.~\ref{fig:FERoneStat} we show the FER of all Rx positions measured for one stationarity region and classified into four classes. We choose $\approx 10\, \text{m}$ distance between the Rx positions. In Fig. \ref{fig:FERoneRx} we show the FER over the Tx trajectory, measured at a specific Rx position during one simulation run.

\subsubsection{Stochastic Channel Model}
In order to obtain more diverse training data we additionally employ a stochastic channel model~\cite{Dak20}. The radio channel is modeled by eight delay taps and an exponentially decreasing power delay profile. Each delay tap contains $40$ propagation paths. They are represented by Rayleigh fading in the non-LOS scenarios, while in the case of the LOS, only the first delay tap is modeled as a Rician fading process. Finally, we obtain additional $8800$ of time-variant frequency response sequences using the statistical parameters from \cite[Table II]{Dak20}.

\subsection{Labels}
\label{seq:def_classes}
\begin{figure}[ht!]
		\includegraphics[width=1\columnwidth]{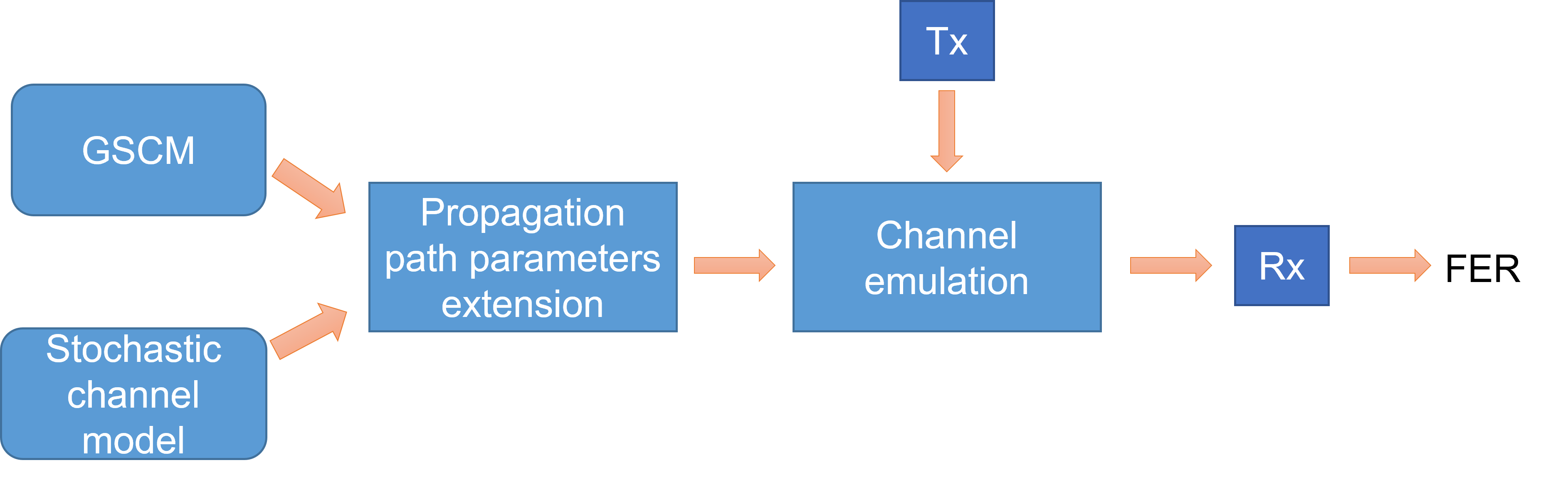}
	\caption{Dataset collection}
	\label{fig:datablock} 
\end{figure}

With the methodology described in the previous chapter we measure the ground truth for the FER. The steps for collecting the ground truth is shown in Fig.~\ref{fig:datablock}. The hardware modems used in this work are the Cohda Wireless MK5 modems~\cite{Cohda}, which implement the IEEE 802.11p standard~\cite{IEEEstandard}. However, we want to emphasize that our methodology is not restricted to a specific hardware nor to a specific communication standard. We use a transmit power of $P_{\text{Tx}}=10\,\text{dBm}$, and a QPSK modulation with a convolutional coding rate of $1/2$. For each stationarity region we emulate $F=20 000$ frames with a size of $100\, \text{bytes}$, using a frame rate of $2200\, \text{frames/s}$. We label our dataset as follows: 
\begin{itemize}
    \item class $1$: $\gamma_{1} := (0, 5\cdot10^{-4}]$,
    \item class $2$: $\gamma_{2} := (5\cdot10^{-4}, 10^{-1}]$,
    \item class $3$: $\gamma_{3} := (10^{-1}, 5\cdot10^{-1}]$, 
    \item class $4$: $\gamma_{4} := (5\cdot10^{-1}, 1]$.
\end{itemize}
The classes are obtained by running a $k-\text{means}$ clustering algorithm on the labeled dataset. 
The output of the DNN model can be interpreted as a discrete probability density on the FER classes. We define the prediction result by the argument of the maximum among all classes. Thus, the prediction result is the FER class having the highest probability.

\section{Deep Neural Network}
\label{sec:DNN} 
The time-varying frequency response of one stationarity region consists of $M=200$ time samples and $N=601$ subcarriers. Since the bandwidth of the Cohda Wireless MK5 modems is $10 \, \text{MHz}$, we have to resample the frequency response for training and hence obtain $41$ frequency samples. Since the frequency response is represented by complex numbers, we split the input dataset into a real and an imaginary part. Hence, one sample of our input dataset is represented by a rank three tensor $\bm{X}_{j,k,l}$, with the dimensions of $(2,200,41)$. By flattening each sample to a one dimensional vector, we finally obtain the input vector $\bm{x}_{P}$, where $P = 2\cdot200\cdot41 = 16400$ represents the number of features of the input vector. The architecture of the DNN, shown in Table~\ref{tab:dnn}, consists of six fully connected linear layers, each followed by a rectified linear unit (ReLU) as an activation function. The DNN is implemented with the PyTorch11 library for Python.

The linear layer can be described as 
\begin{equation}
    \bm{y}_{h} = \bm{W}_{h} \cdot \bm{x}_{h} + \bm{b}_{h},
\end{equation}
where $h$ is the index of the linear layer, $\bm{y}$ represents the output vector, $\bm{W}$ the weighting matrix and $\bm{b}$ the bias. The number of rows and columns of the weighting matrix in the observed linear layer is determined by the output vector size $B_{h}$ and the input vector size $A_{h}$ of this layer, respectively. The size of the bias vector is equal to the size of the output vector. We employ a logarithmic SoftMax function for obtaining a discrete probability density on the FER classes.
\begin{table*}[t]
\begin{tabular}{|l|l|l|}
\hline
\textbf{Layer} &  & \textbf{Number of trainable parameters} \\ \hline
Linear         &  $y_{1}= \bm{W}_{1}x_{1}+ b_{1}, A_{1} = 16200, B_{1} = 2048 $ & $33589248$                               \\ \hline
ReLU           &  $z_\text{1} =  \text{max}(0, y_{1})$  & $0$                                      \\ \hline
Linear         & $y_{2}= \bm{W}_{2}x_{2}+ b_{2}, A_{2} = 2048, B_{2} = 1024$  & $2098176$                                 \\ \hline
ReLU           & $z_{2} =  \text{max}(0, y_{2})$ & $0$                                       \\ \hline
Linear         & $y_{3}= \bm{W}_{3}x_{3}+ b_{3}, A_{3} = 1024, B_{3} = 1024$ & $1049600$                                 \\ \hline
ReLU           & $z_{3} = \text{max}(0, y_{3})$ & $0$                                       \\ \hline
Linear         & $y_{4}= \bm{W}_{4}x_{4}+ b_{4}, A_{4} = 1024, B_{4} = 512$ & $524800$                                  \\ \hline
ReLU           & $z_{4} =  \text{max}(0, y_{4}) $ & $0$                                       \\ \hline
Dropout           & $p=0.05$ & $0$                                       \\ \hline
Linear         & $y_{5}= \bm{W}_{5}x_{5}+ b_{5}, A_{5} = 512, B_{5} = 128$ & $65664$                                   \\ \hline
ReLU           & $z_{5} = \text{max}(0, y_{5})$ & $0$                                       \\ \hline
Linear         & $y_{6}= \bm{W}_{6}x_{6}+ b_{6}, A_{6} = 128, B_{6} = 4$ & $516$                                     \\ \hline
LogSoftMax     & $z_{6} = \log \left(\frac{\exp(y_{6})}{\sum_{j} \exp(y_{6j})}\right)$ & $0$                                       \\ \hline
\end{tabular}
	\caption{Architecture of the deep neural network}
	\label{tab:dnn} 
\end{table*}

Our dataset contains $15916$ samples which we divide it into $70 \, \%$ training and $30 \, \%$ test samples for the model evaluation. We include an equal amount of data from the GSCM and the stochastic channel model in both, the training and the test dataset. During the training process we update the weighting matrix and biases using the Adam optimizer~\cite{Kingma14} with a learning rate of $10^{-4}$. Considering the classification task, the loss function $\rho$, which we calculate, is the cross entropy which reads
\begin{equation}
    \rho(\bm{y},n) = -\bm{y_{\text{n}}} + \ln\left(\sum_{i=1}^{N_{\text{class}}} \me^{y_{i}}\right),
\end{equation}
where $\bm{y}$ represents the observed vector and $n$ the correct class.

\begin{figure}[ht!]

        \begin{center}
        \hspace{0.1cm}
                    \includegraphics[width=0.87\columnwidth]{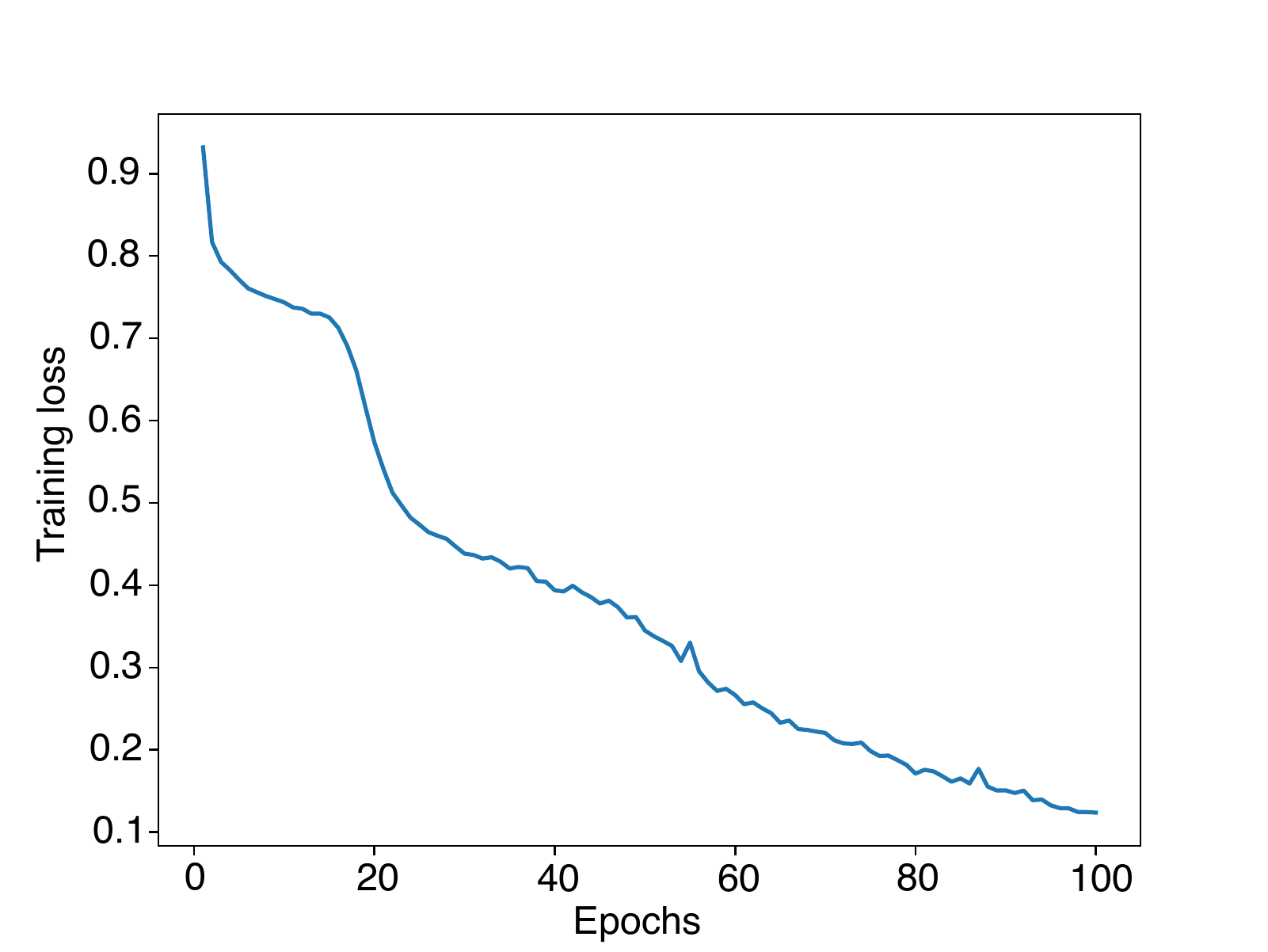}
        \end{center}

	   \caption{History of the training loss.}
	   \label{fig:lossepochs} 
\end{figure}

\begin{figure}[ht!]
    \hspace{-0.2cm}
		\includegraphics[width=0.75\columnwidth]{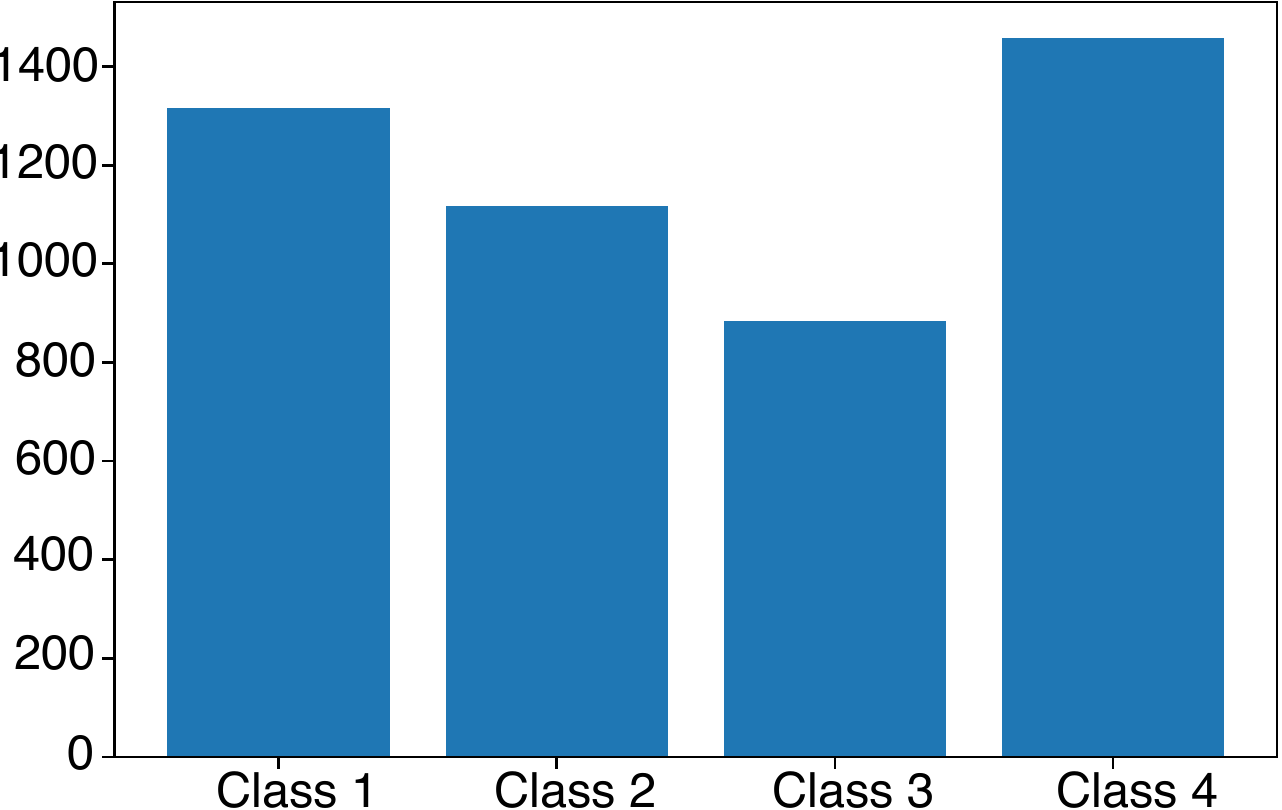}
	\caption{Number of samples per class within the test dataset.}
	\label{fig:clasdistr} 
\end{figure}

\begin{figure}[ht!]
		\includegraphics[width=0.75\columnwidth]{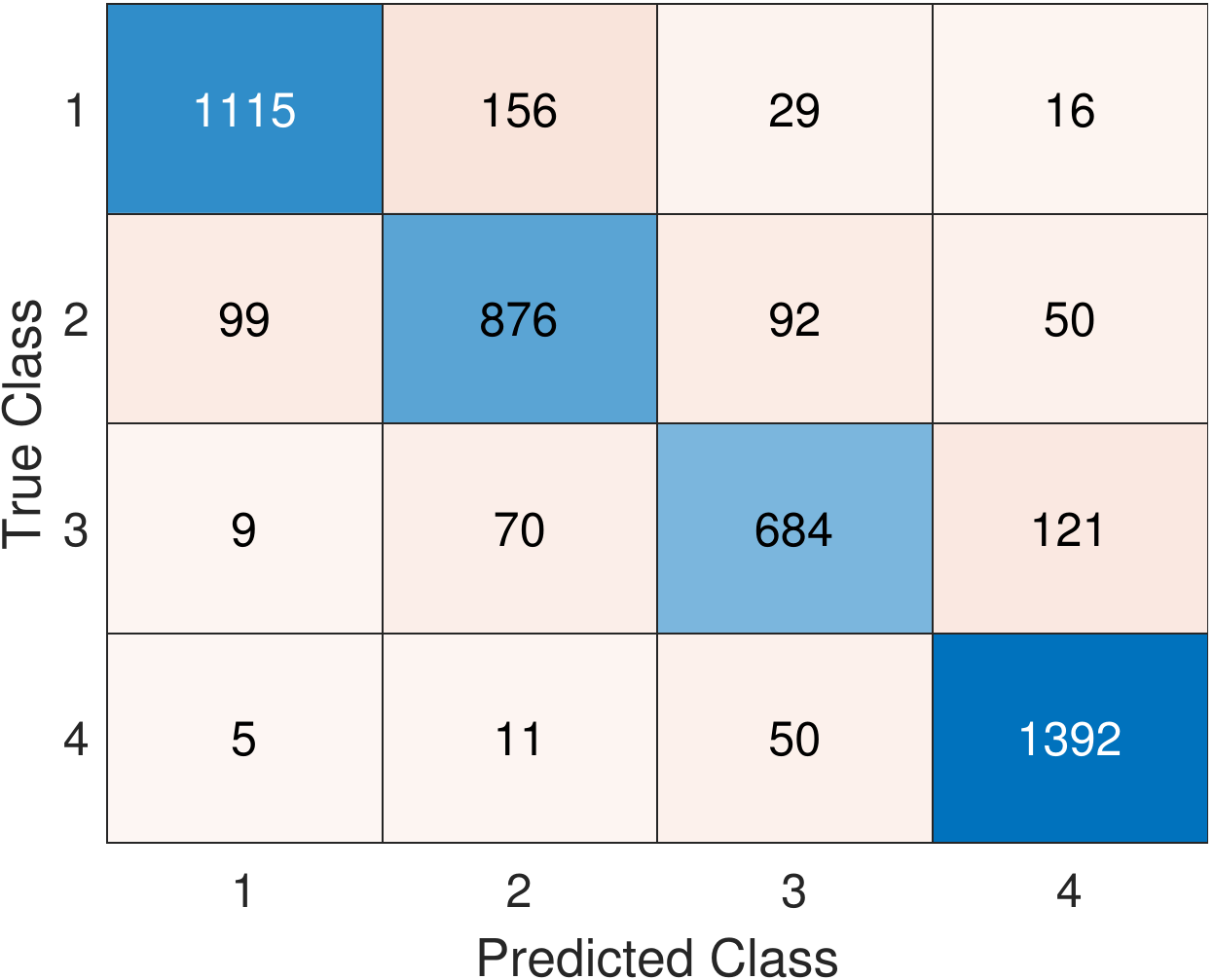}
	\caption{Confusion matrix obtained for the test dataset with a size of $4775$ samples. The diagonal depicts the number of correct  predictions for each class.}
	\label{fig:confmatrix} 
\end{figure}

We train our DNN on a NVIDIA Tesla V100 GPU card with $16$\,GB of RAM. We stop the training when there is only a marginal decrease in loss. This results in a training that takes approximately $100$ epochs which took $15\,\text{min}$ on the mentioned hardware. Fig.~\ref{fig:lossepochs} depicts the loss as a function of training epochs. Furthermore, we evaluate the performance of the trained DNN model on the test dataset with $4775$ samples. Fig.~\ref{fig:clasdistr} shows the class distribution of the test dataset. We present the prediction results using the test dataset in Fig.~\ref{fig:confmatrix}. We obtain a prediction accuracy of $85.2 \,\% $. From the confusion matrix we can see that the best prediction is obtained for the classes containing the lowest FERs (class $1$), with an accuracy of $84.73\,\% $  and containing the highest FERs (class $4$), with an accuracy of $95.47 \,\% $. We can notice that the majority of the samples are in those classes. For the other two classes we obtain an accuracy of $\approx 78 \, \%$.

\section{Validation with Real World Measurements}
\begin{figure}[ht!]
	\begin{center}
    \hspace*{-0.1in}
		\includegraphics[width=0.99\columnwidth]{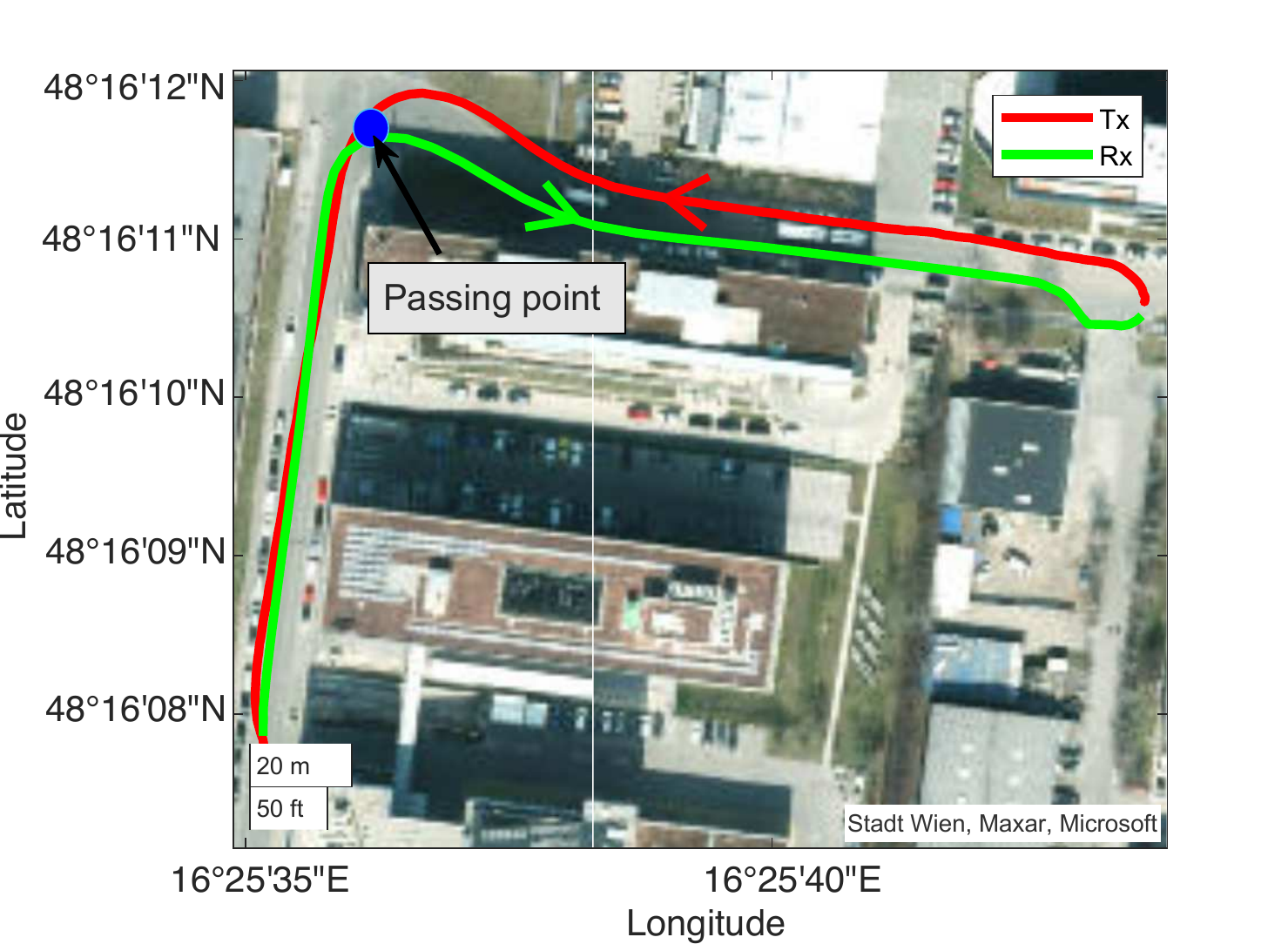}
	\end{center}
	\caption{Scenario used for the validation.}
	\label{fig:scenario} 
\end{figure}

In order to test the applicability of the trained model, we finally predict the FER class for a real-world V2V measurement. The measurement has been conducted in the inner city of Vienna and the measurement data is publicly available~\cite{Rainer22}. The measurement includes in addition to the time-varying channel frequency response, LiDAR, RADAR and FER measurements. The FER in the measurement campaign is collected using the same Cohda Wireless MK5 modems as in the HiL setup and is synchronized to the channel sounder. For the validation of our trained model we use the time-varying frequency responses obtained from the first vehicle-to-vehicle scenario of the measurement campaign, which is shown in Fig.\,\ref{fig:scenario}. A detailed description of the dataset can be found in~\cite{Rainer22}. We want to emphasize that we never used these frequency responses in our training or test dataset. The training and test dataset are exclusively produced by the GSCM and the stochastic channel models as mentioned above. 

\begin{figure}[ht!]
		\includegraphics[width=\textwidth]{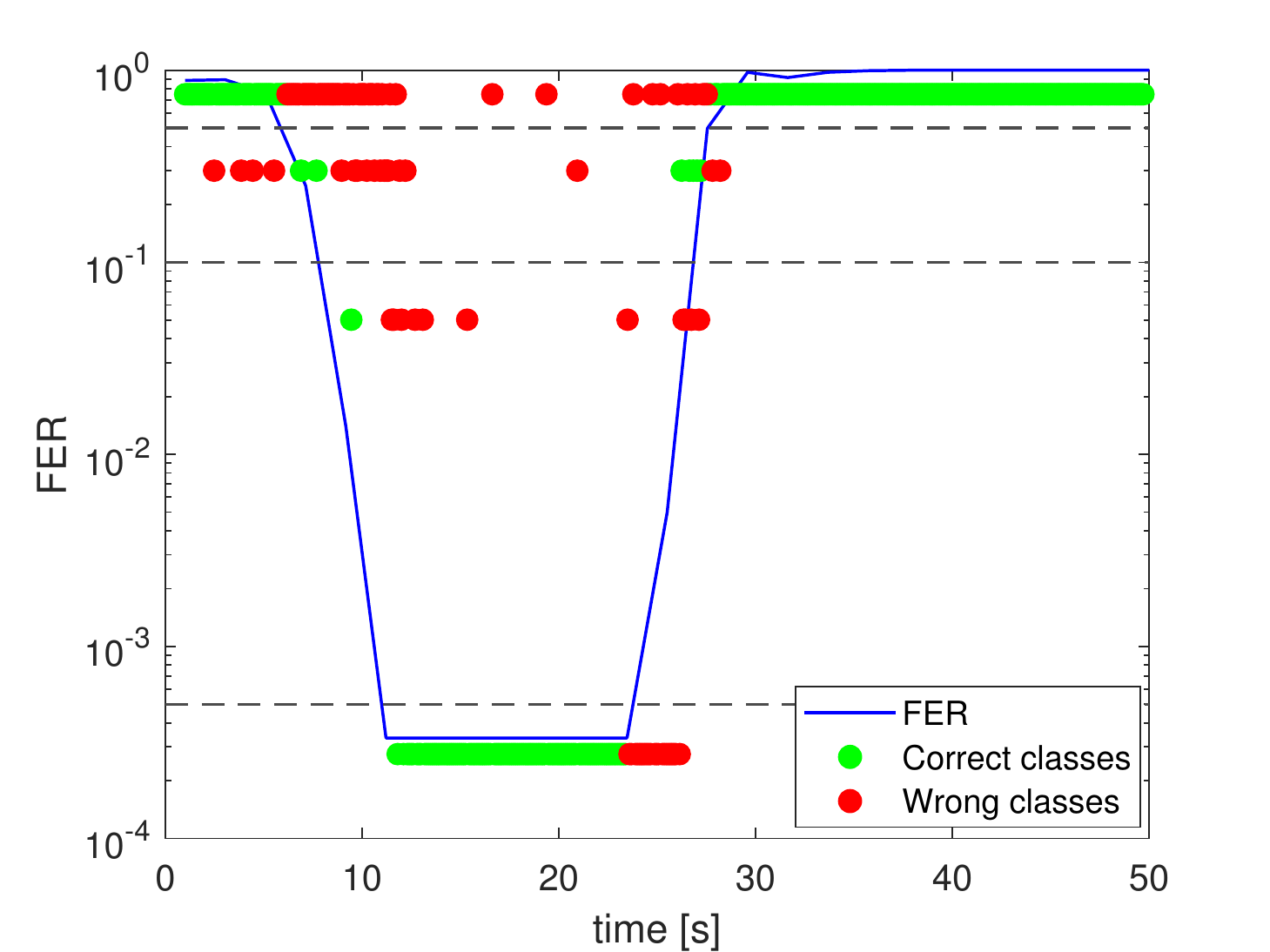}
	\caption{Validation of the DNN with the measured CTFs.}
	\label{fig:validation} 
\end{figure}

The time-varying frequency responses from the measurement campaign are collected by the AIT OFDM multi-band channel sounder~\cite{Zelenbaba20a} with a bandwidth of $150.25\, \text{MHz}$. Therefore we have to adapt the data accordingly to the bandwidth of $10\, \text{MHz}$ such that the trained DNN can process them. Furthermore, we evaluate the FER of the measurement per second and classify them according to the classes defined in Section~\ref{seq:def_classes}. Then, we take the CTF of each stationarity region and give it to the DNN to predict the FER. Fig.~\ref{fig:validation} depicts the measured FER on the road for the selected scenario and whether the trained model predicted the correct FER class. The correct FER class is indicated by a green dot and a red one indicates a false prediction. We notice that most of the samples which belong to the classes with the highest and the lowest FER range are correctly predicted. For the other two classes, class $2$ and class $3$, we get a higher percentage of misclassification, especially between the neighboring classes. Finally, the total accuracy of the prediction with the real-word measured frequency responses is $78\,\%$. These results show that even though the number of transmitted frames per stationarity region is very small, in this case $150$ frames, with our DNN we are able to predict an accurate FER.

\section{Conclusion}
We approached the problem of predicting the FER from raw channel samples by formulating a classification problem with appropriate FER classes. We designed a DNN with six hidden layers that is able to achieve an accuracy of approx. $85\%$ on the labeled dataset. The dataset was only populated by samples obtained by our numerical channel models parameterized by measurements. We provided a methodology to measure the FER of non-stationary vehicular wireless communication links using a specific communication system with arbitrary accuracy. This methodology enables us to overcome the problem of having to average numerous drive-by experiments in order to obtain an expected FER. We used the proposed methodology to obtain the labels (FER) for our dataset using IEEE 802.11p compliant modems. The ability of the trained DNN to generalize is shown by predicting the FER from channel samples obtained by a real-world V2V measurement campaign. The results indicate a good prediction accuracy of $78\,\%$.

\section{Acknowledgments}
This work is funded by the Austrian Research Promotion Agency (FFG) and the Austrian Ministry for Transport, Innovation and Technology (BMK) within the project RELEVANCE (881701) of the funding program transnational projects, by the European Commission within the European
Union’s Horizon 2020 research innovation programme funding
ECSEL Joint Undertaking project AI4CSM under Grant
Agreement No. 101007326 and within the Principal Scientist grant Dependable Wireless 6G Communication Systems (DEDICATE 6G) at the AIT Austrian Institute of Technology.

\bibliography{main.bib}
\bibliographystyle{IEEEtran}

\end{document}